\documentclass{aa}  

\usepackage{amsmath,amssymb}
\usepackage[normalem]{ulem}
\usepackage{bm}
\usepackage{comment}
\usepackage{xcolor}
\usepackage{soul}
\usepackage[percent]{overpic}
\usepackage{overpic}
\usepackage{natbib}

\newcommand{\bma}{\begin{pmatrix}}
\newcommand{\ema}{\end{pmatrix}}

\newcommand{\bea}{\begin{eqnarray}}
\newcommand{\eea}{\end{eqnarray}}

\newcommand{\be}{\begin{equation}}
\newcommand{\ee}{\end{equation}}        
\newcommand{\ba}{\begin{eqnarray}}
\newcommand{\ea}{\end{eqnarray}}

\newcommand{\R}[1]{\textcolor{red}{#1}}

\usepackage[T1]{fontenc}
\usepackage{ae,aecompl}

\begin{document} 

   \title{Pulsar magnetospheric convulsions induced by\\ an external magnetic field}
   \author{Fan Zhang
          \inst{1}\inst{2}
          }
   \institute{Gravitational Wave and Cosmology Laboratory, Department of Astronomy, Beijing Normal University, Beijing 100875, China \\ \email{fnzhang@bnu.edu.cn}
         \and Department of Physics and Astronomy, West Virginia University, PO Box 6315, Morgantown, WV 26506, USA}
   \date{Received July 6, 2016; accepted }

  \abstract
{The canonical pulsar magnetosphere contains a bubble of closed magnetic field lines that is separated from the open lines by current sheets, and different branches of such sheets intersect at a critical line on the light cylinder (LC). The LC is located far away from the neutron star, and the pulsar's intrinsic magnetic field at that location is much weaker than the commonly quoted numbers applicable to the star surface. 
The magnetic field surrounding supermassive black holes that reside in galactic nuclei is of comparable or greater strength. Therefore, when the pulsar travels inside such regions, a non-negligible Lorentz force is experienced by the current sheets, which tends to pull them apart at the critical line. As breakage occurs, instabilities ensue that burst the bubble, allowing closed field lines to snap open and release large amounts of electromagnetic energy, sufficient to power fast radio bursts (FRBs). This process is necessarily associated with an environment of a strong magnetic field and thus might explain the large rotation measures recorded for the FRBs. We sketch a portrait of the process and examine its compatibility with several other salient features of the FRBs. }


   \maketitle
%

\section{Introduction}\label{sec:Intro}
Fast radio bursts (FRBs) are transient radio events with millisecond durations, and they are typically associated with large dispersion measures (DMs). Sources of cosmological and galactic origins have both been proposed (see the review article \cite{2016MPLA...3130013K} and references therein), with the latter challenged by more recent observations and theoretical arguments (\cite{2015Natur.528..523M,2016ApJ...818...19K,2014ApJ...785L..26L}), therefore we assume extragalactic sources to be responsible for
FRBs here.

The enormous distance scales that come with this assumption imply that a large amount of energy needs to be available, $10^{38}-10^{40}$ergs, to be more precise (\cite{2016MPLA...3130013K}, assuming isotropic emission). In addition, the short pulse duration requires a compact source region. Neutron stars (NSs) represent natural candidates that simultaneously satisfy these two requirements. 
The gravitational energy of an NS can obviously serve as a deep energy well, but to draw from it, significant changes (such as a collapse, see \cite{2014A&A...562A.137F}) probably have to occur on the star itself, which would be difficult to repair. This contradicts the recent observation that FRBs repeat (\cite{Spitler:2016dmz,Scholz:2016rpt}), which excludes the possibility that disruptive processes cause irreversible damages to the source. An alternative energy well is the electromagnetic (EM) well residing in the NS magnetosphere. This well is shallower, however, and the FRB energy estimate constitutes a substantial portion of the total EM energy in the magnetosphere of a normal pulsar (magnetic field strength of $10^{11}-10^{13}$G). 
Having this in mind, much attention has been paid to magnetars (see, e.g.,~\cite{2014MNRAS.442L...9L}), which possess a stronger magnetic field (up to $10^{15}$G), for which relatively minor magnetospheric events  would already suffice energetically. 

This thinking underestimates the flexibility of the magnetospheres, however, and the possibility remains that catastrophic global reconfigurations of a normal pulsar's magnetosphere can occur (under the influence of external factors), but without altering the star itself, which then restores the magnetosphere to its usual state. Such a global event produces a large emission region that it takes radio signals milliseconds to traverse (see the beginning of Sec.~\ref{sec:Profile} below), thus providing a more natural explanation for the duration of FRBs, as opposed to having to amalgamate a large number of short-duration events, such as giant pulses of young pulsars (see, e.g., \cite{2015arXiv151109137K} for more discussions). In this paper, we explore this alternative.

To begin with, we note that a large rotation measure has been recorded for an FRB (\cite{2015Natur.528..523M}), suggesting that a strong magnetic field exists near the source. On the other hand, we know that the region surrounding a supermassive black hole (SMBH) is pervaded by a strong magnetic field, which is a vital ingredient in jet launching mechanisms such as the Blandford-Znajek process (\cite{1977MNRAS.179..433B}). Therefore, it is interesting to see if this field can act as the destabilizing external influence that triggers magnetospheric reconfigurations. 

The field strength is sufficient to play this role. The key is to concentrate on a structurally vital place for pulsar magnetospheres called the light cylinder (LC). If the typical pulsar rotation period is $1$ second,
then within geometrized units the angular velocity of the pulsar rotation can be computed as $\Omega=(2\pi/1)(10^6/3\times10^{10})\approx 2\times 10^{-4} R^{-1}_*$. This implies that a particle that corotates with the NS will have to travel at the speed of light if it is located at $\sim 5\times 10^{9}$cm away from the rotation axes of the star. Such a place is termed the LC. 
At such a large distance from the star, the star-generated magnetic field would have dropped to $10^{-11}$ of its strength near the star, assuming a dipolar field out to the LC. 
In contrast, the magnetic field close to a supermassive black hole is limited by the Eddington field strength to $6\times 10^4 M_8^{-1/2}$G (\cite{2008arXiv0810.1055D,2010Sci...329..927P}) where $M_8$ is the mass of the supermassive black hole divided by $10^8 M_{\odot}$.
From active galactic nuclei jet considerations, \cite{1977MNRAS.179..433B} estimated that the field strength must exceed $100$G, while for our quieter Milky Way, the field strength close to the SMBH Sgr A* is about $20$G (\cite{2013pss5.book..243M}). These values are not negligible compared to pulsar's intrinsic field at the LC, and can exert strong influences on the magnetospheric dynamics.  

We begin a rough sketch of this influence in Sect.~\ref{sec:Iso} by computing the current density along current sheets (CSs) that enclose closed (return to the star) magnetic field lines. Even though they are of vital importance (they determine the force-free regions through boundary conditions), a detailed description of the CS dynamical evolution is generally lacking. Nevertheless, we can evaluate (in Sect.~\ref{sec:Force}) the Lorentz force
that they experience when immersed in the background magnetic field near an SMBH, and show that force discontinuities would dismember the bubble enclosure they provide, with the magnetosphere experiencing a catastrophic transition as a result. We then estimate the overall energy released from such a violent event in Sect.~\ref{sec:EnergyEstimate}, and describe in Sect.~\ref{sec:recharge} the process through which bubbles regrow, thereby completing a full repeatable dynamical cycle (noting that FRBs have been observed to repeat). We finally examine some potential complications to our computations in Sect.~\ref{sec:complications}, and evaluate the conformity of the present proposal to salient features of the FRBs in Sect.\ref{sec:Dis}, introducing some new analysis, for instance, on the temporal pulse profile. The results show good agreement with FRB observations, especially with several previously overlooked features of the signals. Finally, we conclude with an outlook for future work in Sect.~\ref{sec:Con}.  

The formulae in this paper are in geometrized units where $c=G=\epsilon_0=1$, unless stated otherwise. With this choice, only one fundamental unit, that of length, is required, and we take it to be the radius $R_*$ of the NS. In other words, a length of $1R_*$ in our formulae corresponds to roughly $10^6$cm in cgs units. The index notation adopted in this paper is that the beginning part of the Latin alphabet denotes four dimensional spacetime quantities, while the middle part of the Latin alphabet denotes spatial components.

\section{Pulsar magnetosphere in isolation} \label{sec:Iso}
Before examining the dynamical evolution brought about by the background magnetic field of the galactic center, we first enumerate some of the important structural features of a pulsar magnetosphere in isolation. We take the outline of \cite{Goldreich:1969sb}, and in particular the quantitative refinement from \cite{2016arXiv160404625G} (GLP for short).
The underlying assumption for their treatment is that the plasma particles' contribution to the total stress-energy tensor of the system is subdominant to that of the EM field, so that the particles' inertia is negligible, and they can experience no forces, or else will be infinitely accelerated. 

GLP also asserted that the spacetime metric outside of the NS ($r>1$ in our units) takes the form of 
\bea
ds^2&=&-\left(1-\frac{\mathcal{C}}{r}\right) dt^2 + \left(1-\frac{\mathcal{C}}{r}\right)^{-2}dr^2 \notag 
\\&&+ r^2 \left[ d\theta^2 + \sin^2\theta\left(d\phi-\frac{\mathcal{C}\mathcal{I} \Omega}{r^3} dt\right)^2\right]\,,
\eea
where $\mathcal{C}$ is a dimensionless compactness parameter with a typical value of $1/2$, and $\mathcal{I}$ is the dimensionless moment of inertia, with a value of $2/5$ for a uniform density sphere. These quantities are dependent on the NS equation of state, and we take the aforementioned example values for concreteness. The quantity $\Omega,$ on the other hand, is the angular frequency of the NS rotation, which we retain as a free parameter in the expressions below. In addition, as the essential ingredients we need already appear for aligned rotators (rotation and magnetic axes of the NS are aligned), 
this is the case we examine. 

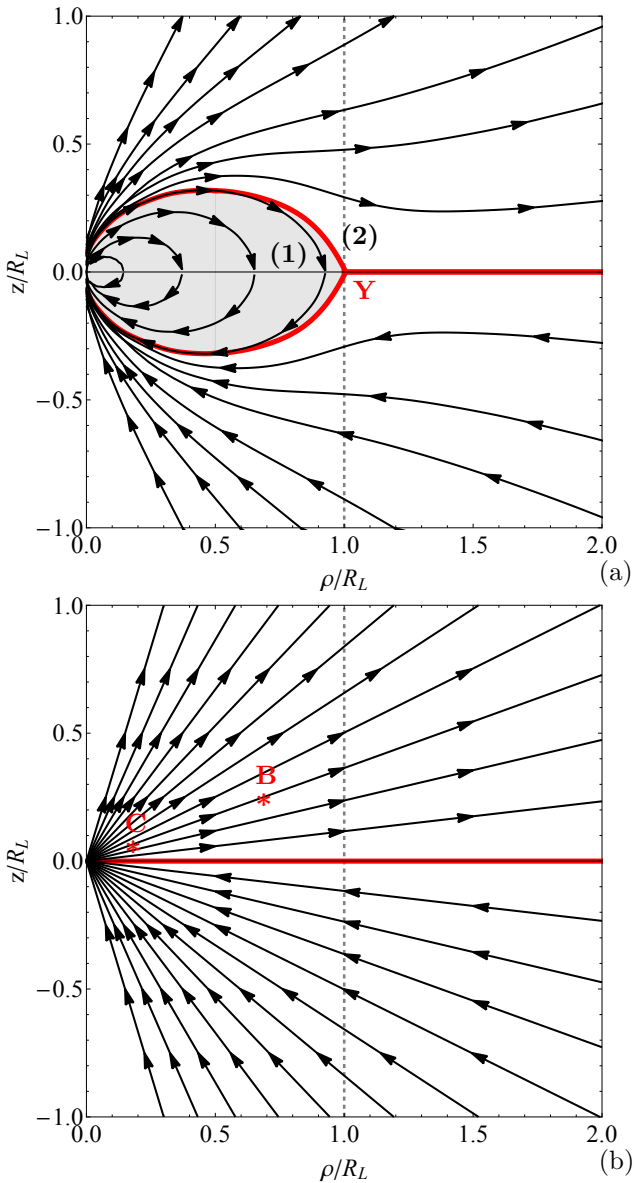
\begin{figure}
\begin{overpic}[width=0.9\columnwidth]{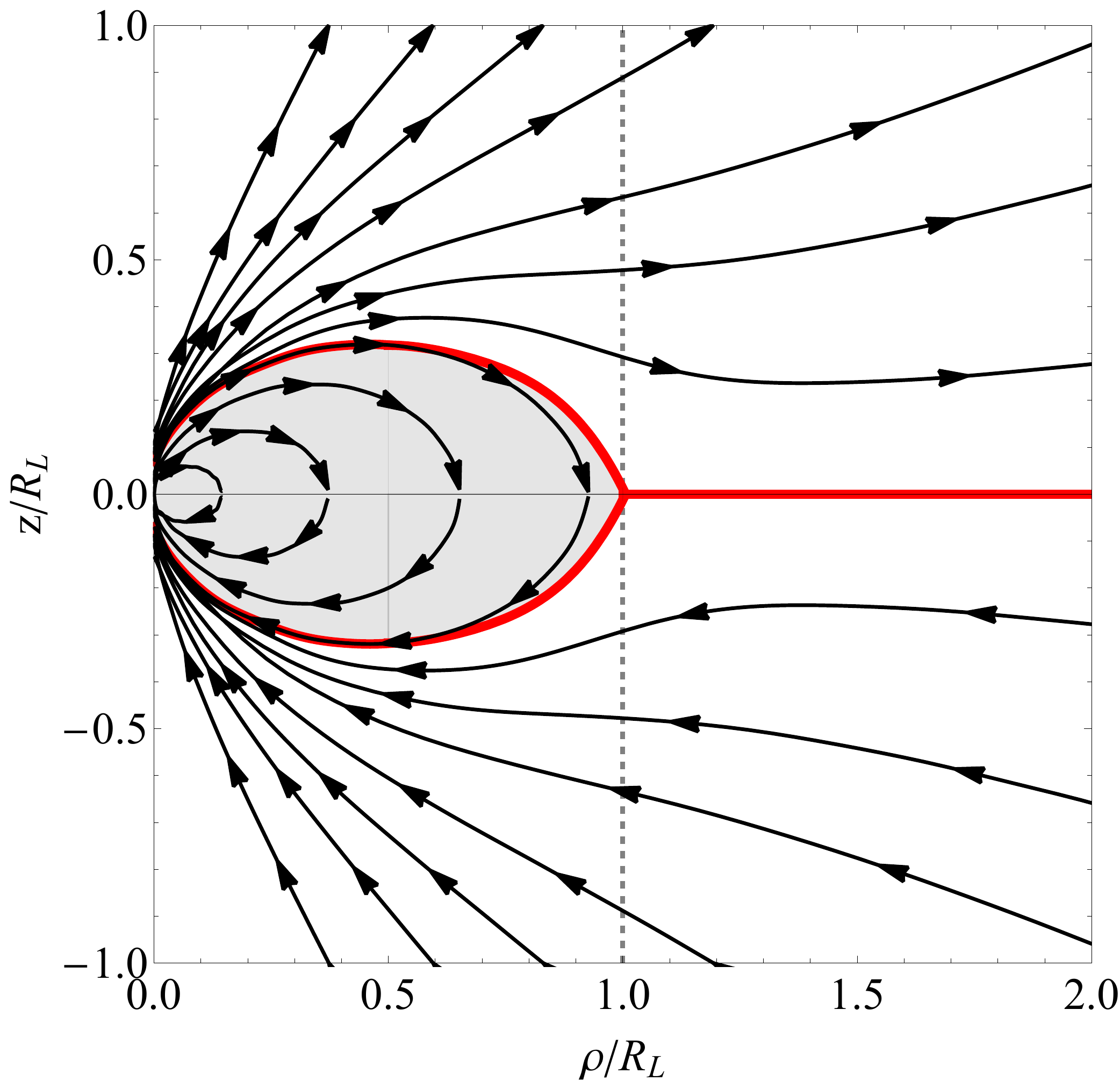}
\put(97,2){(a)}
\put(42,54){{ \bf (1)}}
\put(55,57){{\bf (2)}}
\put(57,48){{\bf \R{Y}}}
\end{overpic}
\begin{overpic}[width=0.9\columnwidth]{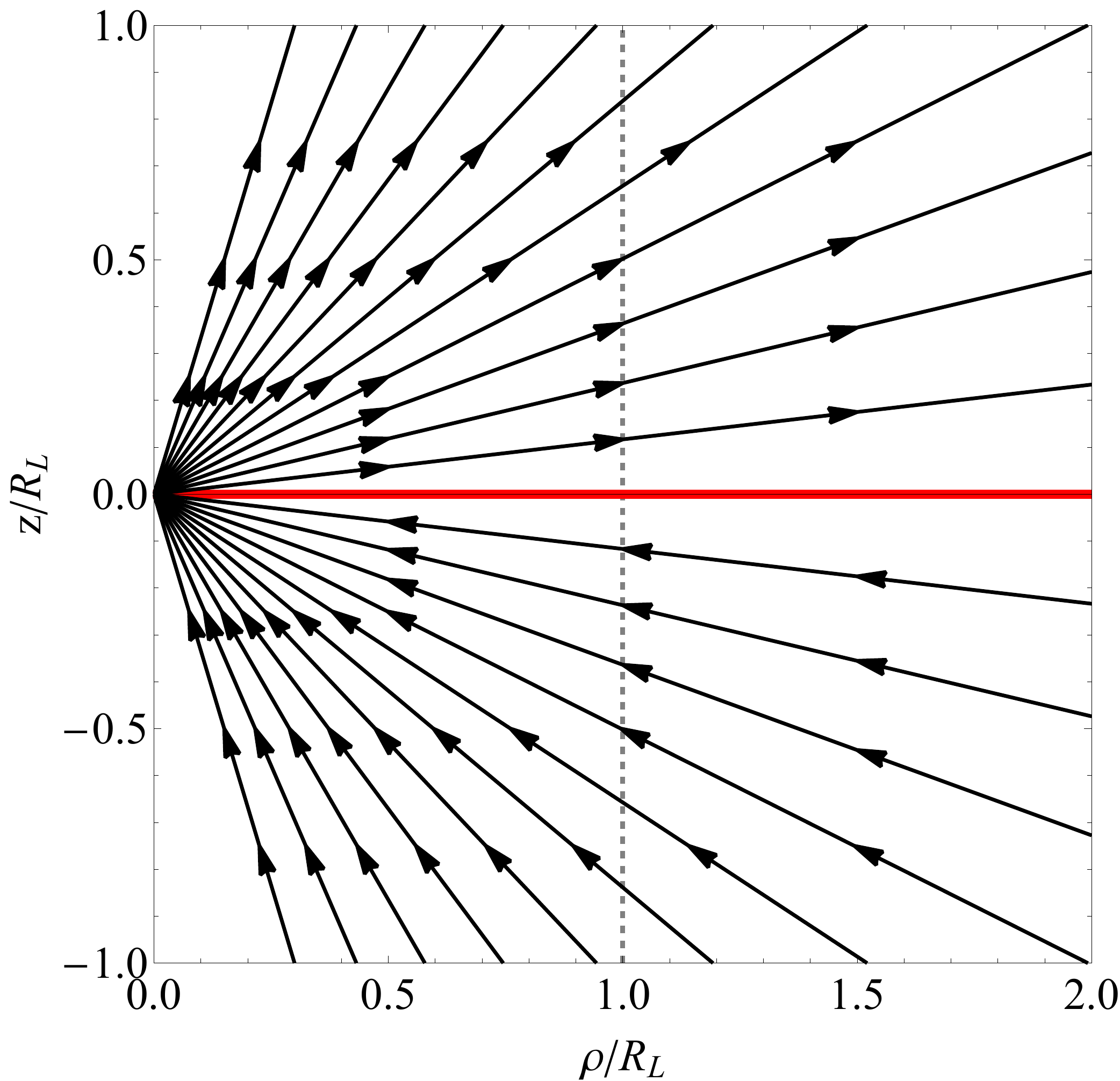}
\put(97,2){(b)}
\put(41.2,60){{\large \bf \R{*}}}
\put(20,52.2){{\large \bf \R{*}}}
\put(41.2,65){{\bf \R{B}}}
\put(20,57.2){{\bf \R{C}}}
\end{overpic}
\caption{(a): Schematic depiction of a poloidal slice of the magnetosphere of an isolated pulsar, reproduced from GLP Fig.~1. The vertical dashed lines mark the location of the LC ($\rho/R_L=1$, with $\rho$ and $z$ being the cylindrical coordinates measuring the distance to and along the rotation axis, and the LC is at $\rho=R_L$). The red curves represent the CSs that demarcate the different magnetic domains $(1)$ (inside the shaded bubble containing closed field lines) and $(2)$ (open field line region).
(b): A depiction of the split monopole solution, with a CS on the equatorial plane, facilitating the change of field line direction across it. 
}
\label{fig:Setup}
\end{figure}

A most salient feature of the pulsar magnetosphere is that a bundle of closed field lines without accompanying currents exists in a bubble, separated from the open field lines outside, along which currents do flow (see Fig.~\ref{fig:Setup} (a)). In order for such regions of distinct characters to coexist, a compressed layer of high current density called the CS needs to be present to separate them (across which the magnetic field is allowed be discontinuous), and the integral version of the Maxwell equations (see Sect.~6 in \cite{Gralla:2014yja}) dictates that the CSs need to be tangential to the magnetic field lines. 

Quantitatively, we note that the Faraday tensor within the \emph{\textup{stationary axisymmetric}} force-free magnetosphere is given by GLP, Eq.~6, in the exterior calculus notation as 
\bea \label{eq:F}
F=\frac{r I(\psi)}{\pi(2r-1)\sin\theta} dr\wedge d\theta + d\psi \wedge\Big( d\phi -\Omega dt \Big)\,.
\eea
The quantity $2\pi\psi(r,\theta)$ is the polar magnetic flux through any surface bounded by the toroidal curve of constant $r$ and $\theta$ (as given by the arguments of the function $\psi$), and $I$ is the polar current through that same surface. The bubble region corresponds to high values $\psi >\psi_0 \approx 1.23 \mu \Omega$, where $\mu$ is the dipole moment of the magnetic field close to the star. In this region, we have $I(\psi)=0$, which is quite different from the outside region with $\psi<\psi_0,$ where (GLP, Eq.~22, $+$ sign for the northern hemisphere)
\bea \label{eq:OpenCurrent}
I(\psi) =\pm 2\pi \Omega \psi \left( 2-\frac{\psi}{\psi_0}-\frac{1}{5}\left(\frac{\psi}{\psi_0}\right)^3\right)\,.
\eea

Although the closed field lines are of secondary importance to pulsars in isolation because they are unrelated to the regular pulsed radiations, they nevertheless serve as a reservoir holding on to energy that is available to be released in a sudden outburst, should the bubble walls experience any catastrophic failure. A general rule for the closed field lines is that they must reside within the LC, never venture outside. The reason being that beyond the LC, the particles that are stuck on the field lines satisfying force-free conditions will have to move superluminally, which is impossible (\cite{Goldreich:1969sb}). Alternatively, magnetic dominance is lost for such field lines beyond the LC (\cite{Gralla:2014yja}). That these two statements are equivalent can be seen simply by writing down the flat spacetime expression for the Lorentz force $q(\bm{E}+\bm{v} \times \bm{B})$, and observe that in an electrically dominated region, we would need $|\bm{v}|>1$ in order to achieve a vanishing force. To see what distiunguishes the closed and open field lines in terms of penetrating the LC, we compute from Eq.~\eqref{eq:F} the invariant (with notation $|B|^2 \equiv B_aB^a$, cf.~\cite{Gralla:2014yja}, Eq.~66)
\bea
F_{ab}F^{ab} &=& 2(|B|^2-|E|^2) \notag \\
&=&\frac{I^2}{r(2r-1)\pi^2 \sin^2\theta} + |d\psi|^2|d\phi-\Omega dt|^2\,,
\eea
and note that $|d\psi|^2 < 0$ as it is space-like and that $d\phi-\Omega dt$ changes from being space-like to time-like when we cross the LC outward bound. Therefore when $I=0,$ as in the case of the closed field lines, the region beyond the LC is an electrically dominated forbidden zone, while for open field lines with non-vanishing $I$, this region can remain magnetically dominated. It is then not surprising that the tip of the bubble enclosing those close field lines is on the LC, as depicted in Fig.~\ref{fig:Setup}(a). 

From Eq.~\eqref{eq:F}, we can also obtain the explicit form of the magnetic field using $B^a =\epsilon^{bcda}F_{bc}\tau_d/2$, where $\epsilon^{abcd}$ is the 4D Levi-Civita tensor, and $\tau_a$ is the time-like one form orthogonal to constant $t$ slices of spacetime. The result is (upper index spatial vector in the basis of $\{\partial_{\phi},\partial_r,\partial_{\theta}\}$)
\bea \label{eq:Mag}
\bm{B} = \frac{\csc \theta\sqrt{2r-1}}{\sqrt{2} r^{5/2}}  
\left\{
\frac{\csc \theta }{\pi} \frac{r}{2r-1}I(\psi), \,\,
\frac{\partial \psi }{\partial \theta},\,\,
-\frac{\partial \psi}{\partial r}
\right\}\,.
\eea
We see that the toroidal magnetic components immediately  inside and outside of the CS have values of 
\bea \label{eq:Btoroidal}
B^{(1)\phi} =0\,, \quad 
B^{(2){\phi}} \approx 
\pm \frac{\sqrt{2}\mu \Omega^2\csc^2\theta }{r^{3/2}\sqrt{2r-1}}
.\eea
Therefore, by the usual boundary condition across the CS ($\bm{\hat{n}}$ is the outward normal to the CS, which is purely poloidal as
a result of axisymmetry)
\bea \label{eq:CurrentDensity}
\epsilon^{ijk} \hat{n}_{j} \left({B}^{(2)}-{B}^{(1)} \right)_k = \sigma^i\,,
\eea
we have that there is a return current (with surface density ${\bm \sigma}$) flowing inward toward the star along the singular separatrix (separating open and closed lines) CS in the poloidal direction. In particular, its distribution is reflection-symmetric against the equatorial plane, or in other words, the currents along the top and bottom red arches in Fig.~\ref{fig:Setup}(a) are both flowing to the left.

\section{Destablizing Lorentz force} \label{sec:Force}
We now place the pulsar in the galactic center, but still far away from the innermost stable circular orbit, so that the general relativistic effects from the SMBH are negligible. The speed of the NS, on the other hand, is approximated by the Keplerian expression $v \sim \sqrt{GM/r}$, which for orbital radius $r \sim 0.005$pc (estimated size of the magnetized region, see Sect.~\ref{sec:Dis} for details) and SMBH mass $M\sim 4\times 10^6{\rm M}_{\odot}$ gives $v \sim 10^{-2}$c (Lorentz factor $\gamma-1 \sim 10^{-5}$), and thus special relativistic effects are not important either. In addition, the electric field strength when we move into the comoving frame of the NS is correspondingly weak. For the purpose of grasping the basics of the proposed FRB mechanism then, it suffices to consider a stationary pulsar immersed in a static magnetic-only background EM field. 

As is clear from Fig.~\ref{fig:Setup}, the CSs form the skeleton of the magnetosphere (it has been observed that force-free solutions are in general numerous and the CS configuration is the deciding factor in selecting particular solutions out of large families of possible candidates (\cite{Goldreich:1969sb,2015PhRvD..91l4055Y})), therefore we concentrate on their reaction to our placing the pulsar in the galactic nuclei. Specifically, they would experience a Lorentz force from the external background field, provided that such fields have not been cancelled by a slight adjustment of the magnetospheric currents. Because the external field strength is comparable to (or easily orders of magnitude stronger than) the star-generated field near the LC, it is clear that small perturbative alterations to the magnetospheric currents are insufficient to shield the background field in that region. The result is that the CSs are moved by the Lorentz force, resulting in a loss of stationarity for the entire magnetosphere, with the CSs serving as moving boundaries to the force-free regions.  

\begin{figure}
\centering
\begin{overpic}[width=0.75\columnwidth]{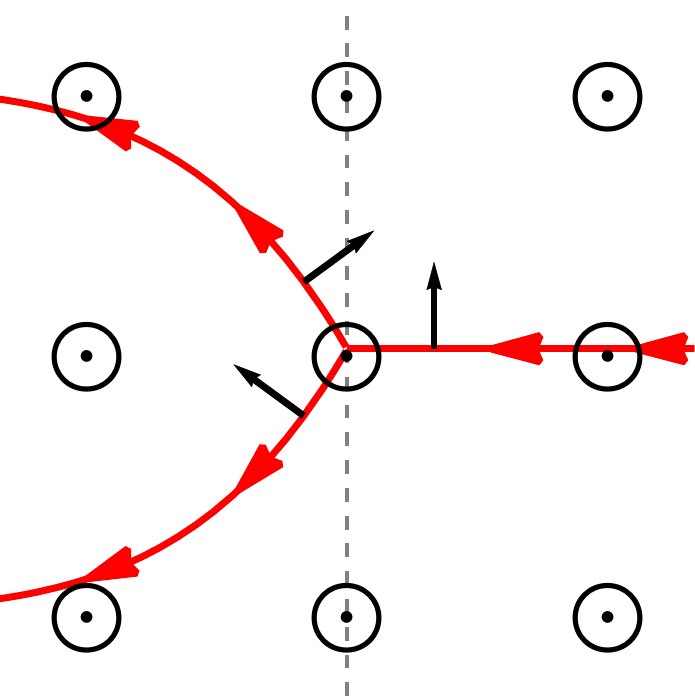}
\put(70,3){(a)}
\put(55,43){{\bf \R{Y}}}
\put(20,44){{\bf B}}
\end{overpic}
\newline
\begin{overpic}[width=0.48\columnwidth]{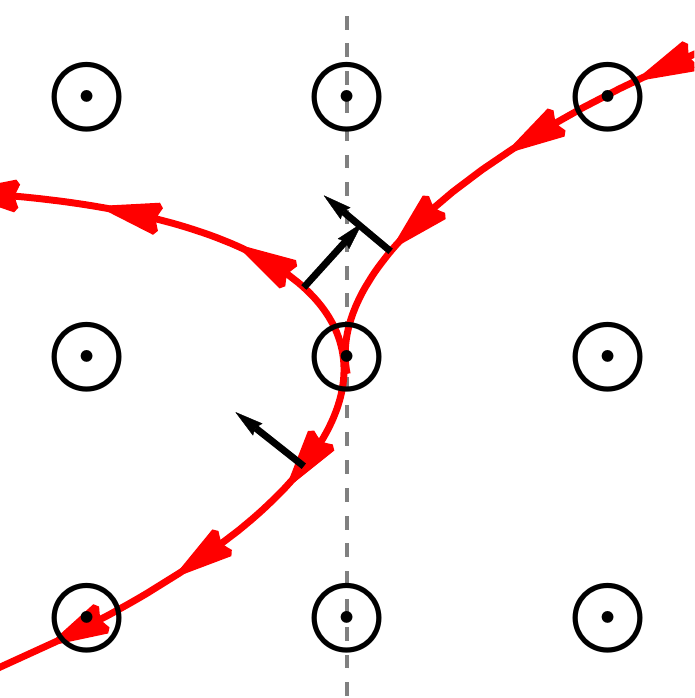}
\put(70,3){(b)}
\put(55,43){{\bf \R{Y}}}
\put(20,44){{\bf B}}
\end{overpic}
\centering
\begin{overpic}[width=0.48\columnwidth]{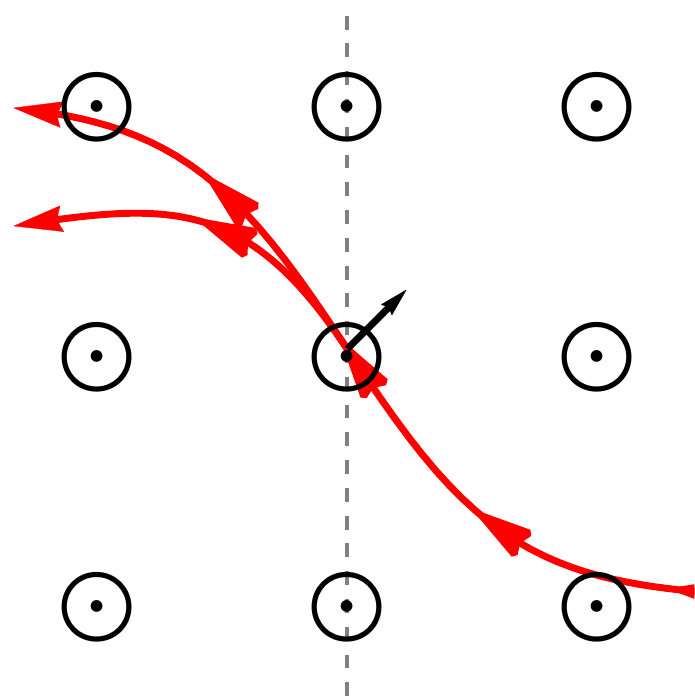}
\put(70,3){(c)}
\put(57,45){{\bf \R{Y}}}
\put(20,44){{\bf B}}
\end{overpic}
\caption{Force distribution near the critical point. The   
background magnetic field points out of the paper, the dotted line denotes the LC, and the black arrows indicate the force directions. (a) The isolated magnetospheric configuration. (b) Slightly adjusted so that the three branches of the current sheets intersect tangentially, but the forces do not synchronize. (c) Heavily modified so that the three branches have matching force directions, but the volume of the bubble is curtailed near the LC.}
\label{fig:Critical}
\end{figure}
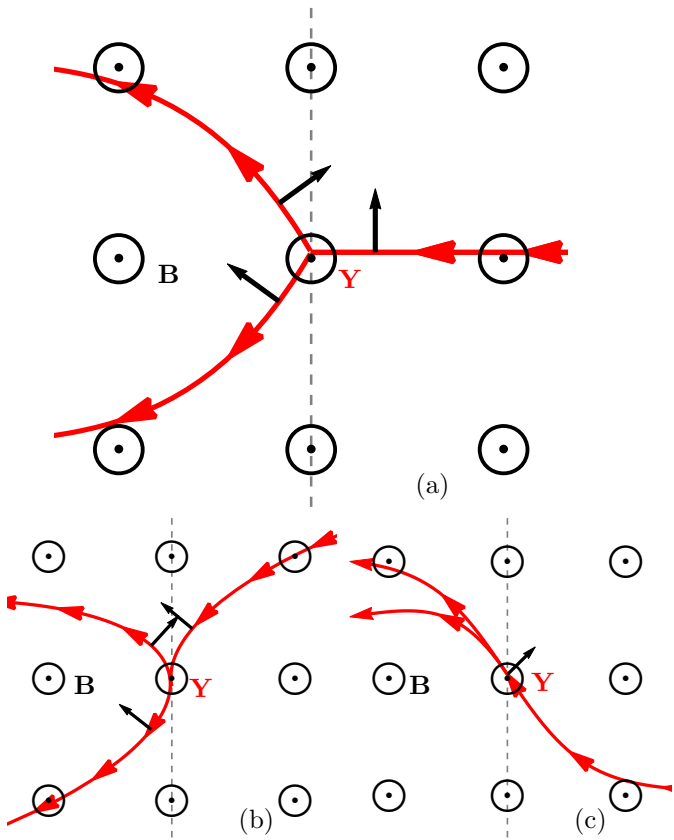

The result of this loss of stationarity is episodic magnetospheric reconfigurations. Specifically, in the regions close to the NS, the CSs in a stationary solution,
if it exists, would be similar to their counterparts in the isolated case with two inflowing currents (computed in Sect.~\ref{sec:Iso}) above and below the equatorial plane with similar fluxes (obeying approximate reflection symmetry) because these regions are still dominated by the magnetic field of the NS itself. However, near the LC, the CS configuration as depicted in Fig.~\ref{fig:Setup} is no longer stable, which means that such stationary solutions cannot in fact exist. This is because the three branches of CSs intersect at a critical point \verb!Y! (see Fig.~\ref{fig:Setup}(a)), where the current flows (with a non-vanishing density ${\bm \sigma} \propto \mu \Omega^2/R_L^2$, see Eqs.~\eqref{eq:Btoroidal} and \eqref{eq:CurrentDensity}) dictate that the different branches are pulled by Lorentz forces in different directions when the non-negligible (near the \verb!Y! point) background magnetic field is orientated as in Fig.~\ref{fig:Critical}(a). In other words, the forces will tear the three branches apart at the seams. While a simple deformation may bring the three branches to intersect tangentially (Fig.~\ref{fig:Critical}(b)), the currents in the top and bottom sheets that flow back to the star are necessarily opposite, so that the forces would still not align at the critical point. A proper force alignment requires more drastic deformations, where the top and bottom branches collapse together at the critical point, as shown in Fig.~\ref{fig:Critical}(c). The force directions will align at \verb!Y!, but the volume of the bubble will be squeezed, and the energy stored in the closed field lines compacted. This implies that significant magnetic pressure ($\propto$ energy density) would have built up inside the bubble that prevented this configuration from being achieved in the first place, especially since the Lorentz force acting on the top separatrix CS also tries to expand instead of compress the bubble. In short, tearing is unavoidable 
\footnote{Beyond tearing, that force acting on the top separatrix sheet even tries to take it across the LC, which would incite further instability, as the closed field lines abutting it and tangential to it will have to accompany it on this journey (our discussion does not rely on this also happening though).}.

The attention is also directed to the fact that the region surrounding the \verb!Y! point is not perfectly force-free; \cite{1994MNRAS.271..621M} predicted that dissipative zones necessarily exist, which means
that magnetic reconnections are allowed there. Therefore, failure in the form of rapidly recurring forced reconnections, and the subsequent disorderly current flows (i.e., charged particle motion), would develop in the CSs at the \verb!Y! point, which is also characterized by high particle speeds and thus a high Reynolds number. The turbulent flows would then be convected along the CSs to regions farther away from the \verb!Y! point, introducing dissipation and resistivity at these places as well. 
Resistive magnetohydrodynamic instabilities such as the rippling and tearing modes (\cite{1963PhFl....6..459F}) would then be able to develop. The configuration with a sudden jump of $B^{\phi}$ across a current layer (see Eq.~\eqref{eq:Btoroidal}) is a stable equilibrium only when the conductivity of the CS is infinite, and once a resistivity $\eta$ is introduced, instabilities grow on a timescale much greater than that of Alfv\'en ($\sim R_L\sqrt{\rho_0}/B^{(2)\phi}$, where $\rho_0$ is the plasma density), but much shorter than that of resistive diffusion ($\sim R^2_L/\eta$), destroying the structural integrity of the CSs and thus removing the segregation between distinct magnetic domains (the magnetic field is no longer allowed to be discontinuous in the absence of CSs, therefore the two domains must assimilate). In other words, the bubble containing the closed field lines that were punctured at the \verb!Y! point would burst open, allowing these lines to spring out into open field lines (to match smoothly with the open field configuration outside), 
along which currents and Alfv\'en waves can travel, possibly as the null solutions of \cite{Brennan:2013jla} (see also \cite{2015PhRvD..92b4049Z} for their stability). When these plasma winds and waves propagate along the originally open field lines near the poles of isolated pulsars, the energy they carry eventually turns into the regular pulsar radio emissions. It is therefore not unreasonable to expect that the same conversion process could channel some of the large amounts of energy initially stored in the closed field lines into a bright, although short-lived, radio burst, or FRB. 

\section{Bubble-free magnetosphere} \label{sec:EnergyEstimate}
The now fully open magnetosphere resembles a split-monopole (see Fig.~\ref{fig:Setup}(b) and \cite{Michel1974,Gralla:2014yja}), which is a valid solution for force-free electrodynamics given by ($q$ being the monopole charge)
\bea \label{eq:Monopole}
F = q \sin\theta d\theta \wedge [d\phi -\Omega d(t-r)]\,,
\eea
and has long been used as a simpler surrogate for the true isolated magnetosphere in regions far from the NS (see, e.g., \cite{Brennan:2013ppa}). A visual comparison between panels (a) and (b) of Fig.~\ref{fig:Setup} shows that the true isolated solution asymptotes to a monopole-like field distribution at large $r$, which is expected because monopole-like radial (in the poloidal directions) field lines is imposed as a boundary condition in GLP. 

The split-monopole is a poor approximate for the isolated magnetosphere when we are closer to the NS, however, as the current inside the star is expected to favor a dipolar near-zone field. However, the field lines of a dipole, given by 
\bea \label{eq:Dipole}
\psi = \frac{\mu}{r} \sin^2\theta\,, 
\eea
are all closed, so that energy cannot be transported out as Poynting or wind fluxes moving out along the field lines. Therefore, to account for the fact that we do observe regular pulsar emissions, field lines near the polar regions will have to be opened up into monopole-like lines by the currents in the magnetosphere. In the regions closer to the equatorial plane, on the other hand, the influence of the dipole is restricted to within the LC, allowing the monopole to dominate outside. 
The bubble is then essentially an island of dipolar dominance in this tug of war, trapping the dipole-like closed field lines within, until released by the collapse of the separatrix CSs, at which time the same split-monopole that dominated outside of the LC becomes dominant everywhere. A turbulent current region close to the star surface (generated during the collapse of the CSs) is then responsible for keeping the influence of the dipole temporarily at bay (allowing for a temporary alteration of the boundary condition near the star). 

Given this general scenario, we can then estimate the energy budget available to be released from the bubble by comparing the energy originally stored there in the form of the dipole solution \eqref{eq:Dipole}, with the same region filled with monopole fields \eqref{eq:Monopole} after the bursting. It is straightforward to compute the energy density, or the ${\cdot}^{tt}$ component of the stress-energy tensor 
\bea
T^{ab}=F^{ac}F^b{}_c - \frac{1}{4} g^{ab} F_{cd}F^{cd}\,,
\eea
which is 
\begin{align} \label{eq:EDensity0}
\rho_E^{l=0}=& \frac{q^2}{50 (2 r-1)^2 r^6} \Big[50 (2 r-1) r^3  \nonumber \\
&+\left(5 \left(5 r \left(8 r^2-4 r+1\right)-8\right) r^3+4\right) \Omega ^2 \sin ^2\theta \Big]\,,
\end{align}
for the split-monopole and 
\begin{align} \label{eq:EDensity1}
\rho_E^{l=1}=&-\frac{\mu ^2}{100 (2 r-1)^2 r^9} \Big[(6 r+1) \cos 2 \theta +10 r-1 \Big] \times \nonumber \\
&\times \Big[\left(1-5 r^3\right)^2 \Omega ^2 \cos 2 \theta-\left(1-5 r^3\right)^2 \Omega ^2+25 (1-2 r) r^3\Big]
\,,
\end{align}
for the dipole (setting $I=0$ as we are interested in the region inside the bubble).
In addition, the magnetic field strengths in the two cases are
\bea \label{eq:MonopoleBStrength}
B^{l=0}=\frac{q}{\sqrt{2} r^2} \sqrt{r (2 r-1) \Omega ^2 \sin ^2\theta +2}\,,
\eea
and 
\bea \label{eq:DipoleBStrength}
B^{l=1}=\frac{\mu\sin \theta }{\sqrt{2} r^4} \sqrt{r \left(8 r \cot ^2\theta +2 r-1\right)}\,.
\eea

Take a pulsar with a magnetic field strength of $\sim B_{11}\times 10^{11}$G at the star surface where the dipole dominates (we note that the variable $B_{11}$ is only the coefficient in front of the power $10^{11}$, and does not include the power term itself, so that for a pulsar with a field strength of $10^{12}$G at the NS surface, we have $B_{11}=10$, and the field strength at the LC for this pulsar would be $1\times B_{11}=10$G), then from Eq.~\eqref{eq:DipoleBStrength} we can work out $\mu$. Since the dipole dominance is taken over by a split-monopole at the LC (at the typical value of $R_L=5\times 10^3 R_*$ for a pulsar of a one-second period), the monopole field strength from Eq.~\eqref{eq:MonopoleBStrength} must be commensurate to the dipole field value at that location, from which we can determine $q$. Because a dipole magnetic field drops off as $1/r^3$ as opposed to the monopole's $1/r^2$, the bubble configuration will have to contain a much stronger magnetic field (as compared to the post-burst monopole) in the near zone to match the same asymptotic field strengths. This is in essence the energy reservoir. Substituting the aforementioned values into Eqs.~\eqref{eq:EDensity0} and \eqref{eq:EDensity1}, we can then integrate the resulting $\Delta \rho_E \equiv \rho_E^{l=1}-\rho_E^{l=0}$ over the bubble region to yield a total released energy of $\approx 4\times 10^{39} B^2_{11}$ergs. This result appears to be in good agreement with the implied energy of $10^{38}-10^{40}$ergs for the FRBs (\cite{2016MPLA...3130013K}), noting that these observation-implied numbers may be slight overestimates, however, depending on which fraction of the DMs is appropriated into the intergalactic medium (see Sect.~\ref{sec:Dis} below). In which case we would have additional room for a much lower radiation efficiency (especially since for a typical pulsar $B_{11} \sim 10$).    

\section{Recharging the bubble} \label{sec:recharge}
Although the split monopole is a valid description of the magnetosphere immediately after the bubble's bursting, the toroidal currents within the NS have not been removed, and they will try to impose a dipolar boundary condition on the magnetosphere (such boundary conditions supported the original dipolar closed field line region in the first place) and restore the closed field line bubble. Immediately after the bursting, their efforts are hindered by a turbulent current layer near the NS surface that temporarily shields the dipolar boundary condition from the magnetosphere. However, the high dissipation within the turbulence would eventually vanquish such currents, and the dipolar influence would begin to push outward again by injecting energy into a nascent bubble, causing it to grow gradually in size. The detailed process most
likely resembles the one outlined in \cite{2007A&A...466..301C}, that is, magnetic reconnection across a resistive equatorial CS (see the red line in Fig.~\ref{fig:Setup}(b)), but applied in a different context inside of the LC: the residual resistivity from the bursting episode inside of the split-monopole's equatorial CS facilitates reconnection of open field lines across the equatorial plane, into closed lines. The process begins near the NS where the toroidal current inside the star bends the field lines into dipolar shapes to be reconnected, and then marches outward. Let $v$ and $\eta$ denote the material velocity and the magnetic diffusivity in the equatorial CS, then the evolution of the closed magnetic field lines along the resistive CS is governed by the induction equation (\cite{2007A&A...466..301C} Eq.~13)
\bea \label{eq:Diffusion}
\frac{dB^i}{dt}=\epsilon^{ijk}\nabla_j\epsilon_k{}^{lm}(v_l - \eta \nabla_l) B_m \,,
\eea
where the first term denotes the advection of the closed field lines by the particle flow in the CS and the second term a diffusion that has a characteristic timescale of $T_{\text{ch}}=R_{L} h/\eta$, where $h$ is the half thickness of the CS. 
\cite{2007A&A...466..301C} integrated and plotted (in their Fig.~2) the outward marching of closed field lines according to Eq.~\eqref{eq:Diffusion} for a toy model. Although that toy model was designed to illustrate the principles for a different process occurring outside of the LC, it turns out to be more directly analogous to our bubble growth process, and we refer to that paper for more details (in particular, their Fig.~2 provides a good visualization for a growing bubble). 
We note, however, that the magnetic diffusivity $\eta$ is proportional to electric resistivity, which means that lower turbulence-induced resistivity in the outer regions (which had more time to settle down before the new bubble reaches them) would prevent effective reconnection of the field lines and thus slow down the growth of the new bubble, causing protracted periods of inactivity and thus low FRB duty cycles. The overall length of the bubble's growth period is otherwise stochastic, as the turbulent post-bursting environment injects variability into both $\eta$ and $v$. Nevertheless, the diffusion timescale $T_{\text{ch}}$ provides a crude estimate for the FRB recurrence interval and thus its repeat rate. We first note that $\eta = 1/(\mu_0 \sigma_0),$ where $\mu_0$ is the vacuum permeability, and $\sigma_0$ is the plasma conductivity, which we approximate by the Coulomb collision formula $\sigma_0 = n_e q_e^2/(m_e \nu_c)$, with $n_e$ being the electron number density, $q_e$ and $m_e$ the electron charge and mass, and $v_c$ the collision frequency. For $n_e$, we can take \cite{Goldreich:1969sb} Eq.~9, which gives 
$n_e = 7\times 10^{-2}B_z \Omega/(2\pi)$, and note that the regrowth process in the outer regions consumes the most time, so that
the relevant $B_z \approx B_{11} \sim \mathcal{O}(10)$G for a typical pulsar (we also adopt a typical $1$s pulsar rotation period). Substituting all these numbers, we obtain $T_{\text{ch}}\approx 10^{-2} h/\nu_c$. Recalling that FRB 121102 has been observed to repeat on $10$min intervals (\cite{Spitler:2016dmz}), we obtain $\nu_e \approx 2\times 10^{-5} h$ $\text{s}^{-1}$, which translates into an effective thermal motion temperature of $\sim 1.4\times 10^4/h^{2/3}$K, broadly in line with temperatures typically found near pulsars (e.g., \cite{1996rftu.proc..173P} and \cite{2001ApJ...552L.129P} found a surface temperature of $10^6$K for the Vela pulsar).

During the growing phase, while the bubble remains small and hidden inside regions of strong intrinsic pulsar magnetic field, the external field can be shielded by perturbations to the intrinsic field, and in this way, the bubble is protected against the destabilizing effects described in Sect.~\ref{sec:Force}. When its \verb!Y! point once again breaches a certain threshold radius and reaches close to the LC, however (i.e., when the lost energy has been replenished by the NS), the intrinsic pulsar magnetic field becomes subdominant in strength to the external field and will not be able to neutralize the latter through its own perturbations. Consequently, Lorentz-force-induced instabilities set in once again to burst the newly grown bubble, and the cycle repeats itself. When the environmental field strength is denoted as $B_{e}$G, the stability threshold is approximately located near $(B_{11}\times 10^{11}/B_e)^{1/3}R_*$, where the intrinsic and external magnetic fields are of comparable strengths. As the \verb!Y! point will not push beyond the LC, we need this threshold radius to be inside of the LC for any instability to occur at all, which gives us a criterion (an additional criterion is to be given in Eq.~\eqref{eq:BeBound} below)
\bea \label{eq:BeBound1}
B_e \geq (\Omega R_*)^3 B_{11}\times 10^{11} \text{G},
\eea 
meaning that the pulsar has to be sufficiently close to the SMBH such that $B_e$ is large enough to trigger our FRB mechanism. We further note that the short bursts we propose are not induced by sudden changes in the environmental magnetic field, but rather are a violent constituent episode in an intrinsic dynamical cycle for the pulsar magnetosphere immersed in a steady external field. In other words, the episodic short-duration FRBs is due to the intrinsic cyclic dynamics of an out-of-equilibrium nonlinear system, and does not require an external trigger with a similar temporal variation profile, somewhat like the crashes in business cycles that do not need to be induced by wars or natural disasters. 

\section{Complications} \label{sec:complications}
\subsection{Tidal forces} \label{sec:tidal}
We have placed the pulsars close to the SMBH, which means that tidal forces might become significant. In this section, we briefly estimate their strengths as compared to the Lorentz forces to obtain a rough idea of when tidal effects have to be taken into account in our computations.

To evaluate the tidal forces, we note that the tide-induced relative acceleration $\Delta a$ between two freely falling observers that are spatially separated by a vector $\xi$ is given by (see \cite{2011PhRvD..84l4014N}) 
\bea
\Delta a^j = -\mathcal{E}^j{}_k \xi^k\,,
\eea
where $\mathcal{E}$ represents the tidal tensor. For a Kerr black hole of mass $M$ and dimensionless spin $a$, the tidal tensor as measured by locally nonrotating Boyer-Lindquist observers is given by (see \cite{2012PhRvD..86h4049Z})
\bea \label{eq:TidalTensor}
\mathcal{E}_{ij} = \bma 
-Q_e \frac{2+\eta}{1-\eta}& \mu Q_m& 0\\
\mu Q_m& Q_e \frac{1+2\eta}{1-\eta}& 0\\
0&0&Q_e 
\ema\,, 
\eea
whereby 
\begin{align}
Q_e &= \frac{Mr (r^2-3a^2 \cos^2\theta)}{\Sigma^3}, \notag \\
Q_m &= \frac{M a \cos\theta (3 r^2 -a^2 \cos^2\theta)}{\Sigma^3},  \notag \\
\eta &= \frac{\Delta a^2\sin^2\theta }{(r^2+a^2)^2}\,, \quad \mu = \frac{3\sqrt{\eta}}{1-\eta}\,, \notag \\
\Sigma &= r^2 + a^2 \cos^2 \theta\,, \quad 
\Delta = r^2 -2 M r +a^2\,,
\end{align}
and $(r,\theta)$ are Boyer-Lindquist coordinates. To represent the directional dependence of the tidal forces in a more explicit and economic manner (as a result of frame dragging, the maximum tidal effect does not necessarily manifest itself in the radial direction), we can compute the eigenvalues of the matrix \eqref{eq:TidalTensor}, which are
\begin{align}
\lambda_1 &= -\frac{Q_e}{2}-\sqrt{\left(\frac{3Q_e}{2}\right)^2\left(\frac{1+\eta}{1-\eta}\right)^2+\mu^2 Q^2_m}\,,\notag \\
\lambda_2 &= -\frac{Q_e}{2}+\sqrt{\left(\frac{3Q_e}{2}\right)^2\left(\frac{1+\eta}{1-\eta}\right)^2+\mu^2 Q^2_m}\,,\notag \\
\lambda_3 &= Q_e\,.
\end{align}
These eigenvalues correspond to the tidal field strength in the three principal (eigenvector) directions of the tidal tensor. 

\begin{figure}
\begin{overpic}[width=0.99\columnwidth]{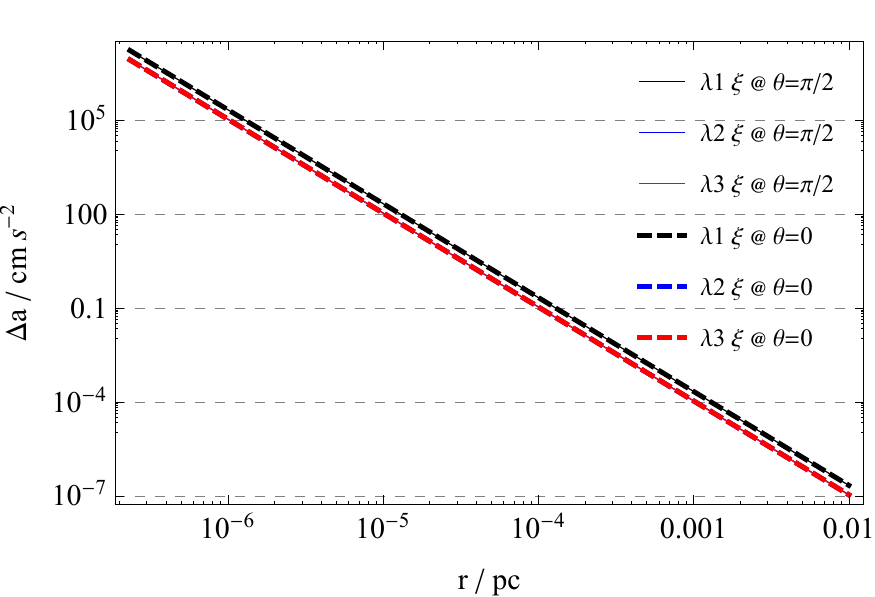}
\end{overpic}
\caption{Relative accelerations corresponding to the three principal directions of the tidal tensor (the blue curves are covered by the red ones). Both axes are in log scale.
}
\label{fig:Tidal}
\end{figure}

Using Sgr A* as a concrete example ($4.9$ million solar masses and a spin of $0.996$) for the SMBH and $|\xi| \approx 5\times 10^9$cm (distance from the star to the LC) as a generous upper limit for the length scale inside the LC, we can compute the tidally induced relative acceleration in the three principal directions as a function of the distance $r$ from the pulsar to the SMBH. We plot the results for two angles $\theta =\pi/2$ and $0$ (i.e., when the pulsar is on the equatorial plane and in the polar regions of the SMBH) in Fig.~\ref{fig:Tidal}, for $r$ ranging from the black hole event horizon to about $0.01$pc, which has been used in the literature as a rough bound for the galactic center region (see references in Sect.~\ref{sec:EventRate}). We can compare this tidal acceleration with that resulting from the Lorentz force, using $B=20$G as an example (applicable for our Milky Way, see Sect.~\ref{sec:Intro}), and noting that the closed magnetic field lines at the LC are loaded with electrons and positrons traveling close to the speed of light, we arrive at a Lorentz-force-induced acceleration on the order of $10^{19}\text{cm s}^{-2}$. A close examination of Fig.~\ref{fig:Tidal} then shows that the tidal effects can be safely ignored all the way down to the event horizon. The reason for this is of course that the black hole is very large, so that although the resulting gravitational field is strong, it varies slowly on a very long length scale. 

\subsection{Pulsar wind}
In Sect. \ref{sec:Iso} we have sketched an analytical outline for the structure of pulsar magnetospheres in regions extending as far out as the LC. In real astronomical surroundings, relativistic winds typically dominate in regions farther out, forming a termination shock front at a stand-off distance $R_s$ 
where the wind pressure balances that of the surrounding medium. This stand-off distance for a bow shock is expected to satisfy (see, e.g., \cite{2006ARA&A..44...17G})
\bea \label{eq:Standoff}
\frac{L}{4\pi R_S^2 c} =\rho_0 v_P^2\,,
\eea
where $L$ is the spin-down luminosity and $v_P$ the speed of the pulsar, while $\rho_0$ is the density of the interstellar medium. Substituting some typical values for field pulsars listed in \cite{2006ARA&A..44...17G}, which are $L \approx 10^{33} \text{ergs s}^{-1}$, $v_P\approx 5\times 10^7 \text{cm s}^{-1}$, and $\rho_0 \approx 1.67\times 10^{-25} \text{g cm}^{-3}$ (hydrogen with a number density of $0.1$ per $\text{cm}^{3}$), we obtain $R_S \approx 2.5\times 10^{15}\text{cm}$, which is far greater than the LC radius of $\approx 5\times 10^9 \text{cm}$. Although both $\rho_0$ and $v_P$ should statistically be greater for pulsars in the galactic center than their field counterparts, the vast gap in order of magnitudes from our estimate suggests that $R_S$ would still most likely to be of sufficient size to enclose the closed field line region, forming in effect a cocoon that isolates this region from the outside environment, at least in terms of matter contents. It is therefore important to assess whether an effective shielding from the magnetic field near the SMBH is also thus established (in addition to the interstellar particles being blown away), which may shut down our FRB mechanism.

Our conclusion is in the negative, since in our strong field scenario, the blowing away of interstellar charged particles by the wind does not translate into an expulsion of the environmental magnetic field. To see this, we note that the magnetic pressure from a $B_e=20$G galactic center magnetic field (we note that the pulsar wind literature is typically concerned with much weaker fields, e.g., the Crab Nebula has an estimated field strength of $300\mu$G (\cite{1982RvMP...54.1183T}), while 3C 58 carries $80\mu$G (\cite{1992MNRAS.258..833G})) is around $10^7 \text{MeV cm}^{-3}$, which when equated with the left-hand side of Eq.~\eqref{eq:Standoff}, which represents the pulsar wind pressure, yields the stand-off radius against the magnetic pressure at $\approx 10^{10}$cm. Comparing this with the LC radius of $\approx 5\times 10^9$cm for a pulsar of period $1$s, we see that a slight increase in the period and/or decrease in the spin-down luminosity and/or increase in the $B_e$ will allow the environmental magnetic field to penetrate the closed field line region and trigger the instabilities described in Sect.~\ref{sec:Force}. The pulsar wind consideration thus sets a (an astronomically achievable) lower bound on $B_e$ that has to be satisfied in addition to Eq.~\eqref{eq:BeBound1}, so that now we have an overall threshold of
\bea \label{eq:BeBound}
B_e \geq \text{max}\left\{(\Omega R_*)^3 B_{11}\times 10^{11} \,, \sqrt{\frac{2 L}{c}}\Omega\right\} \text{G}\,.
\eea

\section{Observables} \label{sec:Dis}
We have proposed a candidate mechanism for the FRBs, which relies on the interaction between the pulsar magnetospheres and a strong background magnetic field within galactic nuclei. This scenario needs to comply with the observed properties of the FRBs. Some have already been discussed; we consider the rest in turn.
  
\subsection{Intrinsic pulse duration and temporal profile} \label{sec:Profile}
A distinguishing feature of the present model is that the release of energy locked up in the bubble is a global event, whose observation signatures are intrinsically of millisecond durations.  
Specifically, the released energy streams out along sprung-open magnetic field lines in an orderly fashion, 
so that only those source locations along field lines that point toward Earth will contribute to the signals we observe. Furthermore, radially along such lines, 
only those regions of sufficiently high energy density (i.e., close to the star) will contribute fluxes that register above
the background noise. When we assume that such regions extend to $\sim 10-100 R_*$, it will take $\sim 0.3-3$ milliseconds for lights to traverse (Aflv\'en waves in force-free plasma also travel with a group velocity equaling the speed of light). This is in contrast to other pulsar events such as giant pulses, which last microseconds. This would help explain the rather broad FRB pulses, especially since evidence of multipath scattering is sometimes lacking (\cite{Spitler:2016dmz}).  

\begin{figure}
\begin{overpic}[width=0.99\columnwidth]{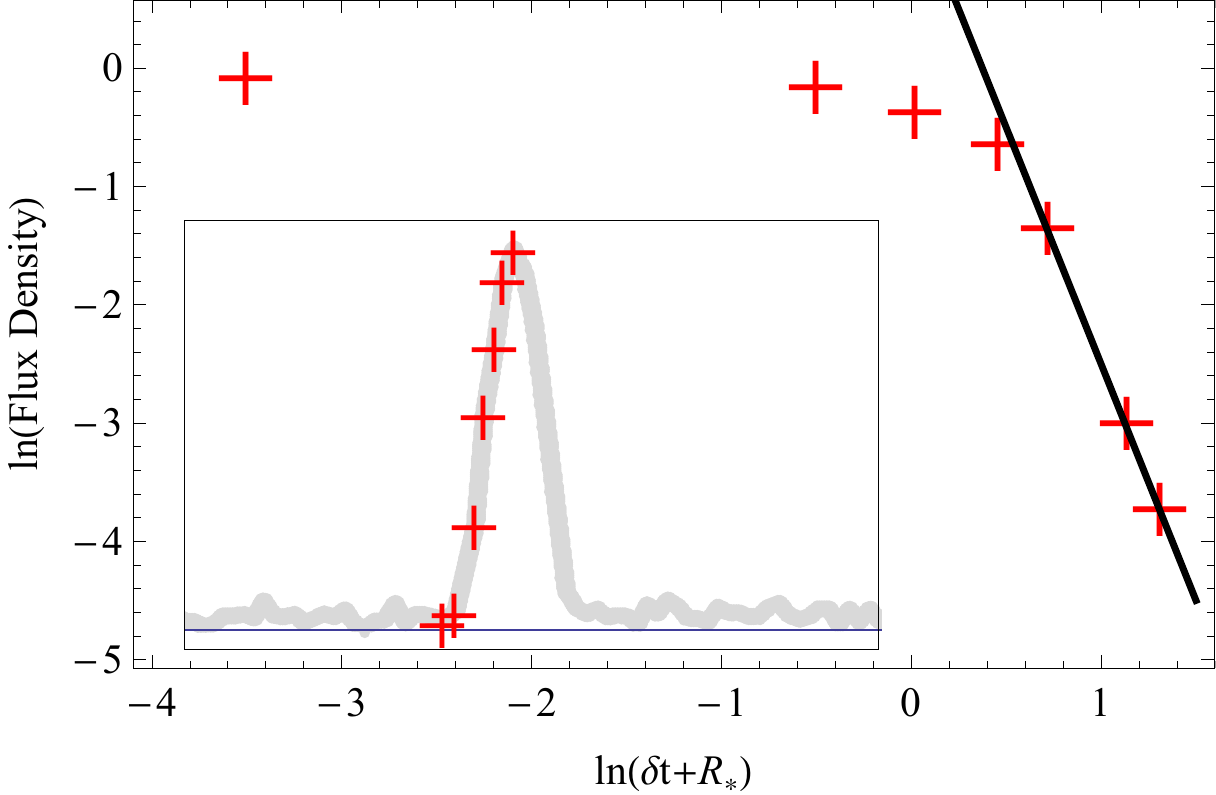}
\put(97,2){(a)}
\end{overpic}
\begin{overpic}[width=0.99\columnwidth]{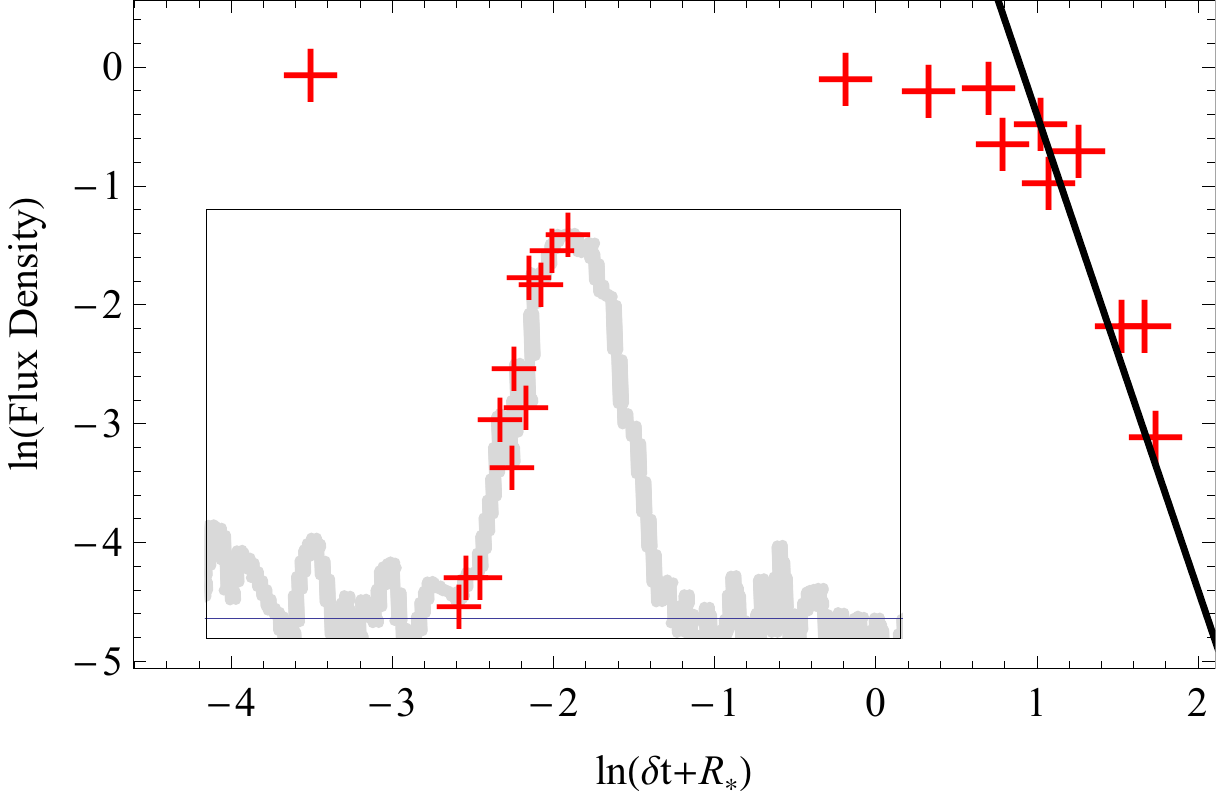}
\put(97,2){(b)}
\end{overpic}
\caption{Log of flux (in arbitrary units) against the log of time to peak-amplitude, with a reference straight line (black) $-4 \ln(\delta t+R_*)+\mathcal{A}$. The inset indicates the location of the data points on the overall signal profile (with space between data points filled in with straight lines). The horizontal line in the inset is the zero flux. 
(a) Burst 11 of the repeating FRB 121102. 
(b) The Lorimer burst FRB 010724. 
}
\label{fig:Fitting}
\end{figure}

We can even predict more than just the overall pulse width. The energy originally stored farther away from the star (e.g., at point \verb!B! in Fig.~\ref{fig:Setup}(b) as compared to point \verb!C!) would have a head-start in terms of streaming along the field lines pointed toward Earth and will arrive at our radio telescopes slightly earlier. Because the density of the released energy
is lower at \verb!B! than it is at \verb!C!, we expect to see an increase in the FRB signal at its leading edge, beginning a few hundred milliseconds before the peak time, but only becoming strong enough to rise above noises much later. Semi-quantitatively, the released energy is roughly the difference between a dipole ($\rho_E^{l=1} \sim r^{-6}$) and a monopole ($\rho_E^{l=0} \sim r^{-4}$). On the other hand, the transverse area subtending the same solid angle is $\propto r^2$, which means that the energy density along the radial direction is $\propto r^2 \Delta \rho_E$.  
Near the peak of the signal (arriving at $t_p$) where energy is released from close to the star, the monopolar term is subdominant and can be dropped. We have then that the log of the flux is approximately 
\bea \label{eq:LeadingEdge}
-4\ln(\delta t +\mathcal{R})+\mathcal{A}\,,
\eea
where we have used $r-\mathcal{R}\approx t_P-t =\delta t,$ with $t_P$ being the arrival time for the peak of the signal and $\mathcal{R}$ the radius from which the peak radiation comes, which we set to $\mathcal{R}\approx R_* =10^6$cm or $0.03$ms in temporal units. We can then fit the logarithm of the frequency-summed burst profiles to expression \eqref{eq:LeadingEdge} by varying the overall amplitude $\mathcal{A}$. 

Because $0.03  $ ms is a small offset, the pulse profile is expected to be rather steep near the peak (without it, the index $-4$ power law would diverge there), so that given a fixed temporal sampling interval, only few data points would land on the leading edge. Therefore we should look for the strongest signals for
which more of the pre-peak segments rise above the background noise. The cleanest and simplest (single-peaked) signal with a high signal-to-noise ratio and several data points on the leading edge is burst 11 reported in \cite{Spitler:2016dmz}. 
We display it in Fig.~\ref{fig:Fitting} (a), and observe that the early part of the leading edge appears to agree with a power law with index $-4$, but the part closer to the peak deviates from it. This may be due to other pulse-smearing effects experienced during signal propagation, or if intrinsic, possibly indicates the existence of complicated dynamics such as turbulences very close to the star that prevent an efficient transportation of the released energy. 
Another strong burst that had multiple data points on the leading edge of its pulse profile is the Lorimer burst (\cite{2007Sci...318..777L}), which, as seen from Fig.~\ref{fig:Fitting}(b), displays a similar morphology.  
For prudence, we caution that several points at the very beginning of the pulses are not above the larger noise peaks during quiescent periods, and that the precise slope of the fitting lines will be influenced by DC offsets in the flux (i.e., the uncertainty in where true zero is). There is thus the need for more dedicated high temporal resolution searches. 

\subsection{Rotation measure} 
A close examination of FRB 110523 yields a rotation measure (RM) of $\sim -186$ rad $\rm m^{-2}$, far in excess of that to be expected from the intergalactic medium and the signal's journey through the Milky Way, which when combined, only gives an RM that is an order of magnitude lower (\cite{2015Natur.528..523M}). The present proposal places the source pulsar within a strongly magnetized region (SMR) in the galactic nucleus, and is therefore endowed with the ability to account for the dominant remainder. The expression for computing the RM is 
\bea
\frac{\rm RM}{\rm rad\,  m^{-2}} = 0.812  \int \frac{n_e}{\rm cm^{-3}}\frac{B_{\parallel}}{\rm \mu G}\frac{dl}{\rm pc}\,,
\eea
where $B_{\parallel}$ is the magnetic field along the line of sight, and $n_e$ is the electron density. Ideally, we would like to know all three terms on the right-hand side and see if we can reproduce the observed RM. However, we know very little about the status of the interstellar medium in the close vicinity of a SMBH, and thus $n_e$ is uncertain. As a surrogate test, then, we attempt to paint a general picture for $n_e$ within the present proposal instead, and see if its inferred value from the RM is at least broadly consistent with this picture. 

We first note that the SMR is expected to be relatively clean outside of the accretion disk (we are statistically unlikely to be viewing the host galaxy disk exactly edge-on, therefore
the FRB signal path to Earth traverses mostly non-disk parts of the SMR, regardless of whether the source pulsar is itself embedded in the disk),
as compared to farther out in the galactic nucleus. This is because charged particles need a source of resistivity to cross the magnetic field lines and therefore are prevented from entering our region of strong magnetization (except for those in the accretion disk near the equatorial plane where mechanisms for strong Ohmic dissipation exist, see \cite{1977MNRAS.179..433B}). Magnetic reconnection may also generate heat and buoyancy in infalling gas, pushing it out (\cite{2002ApJ...566..137I}). 
We therefore expect magnetic dominance (i.e., matter stress-energy density is negligible as compared to that of the EM field), and thus a force-free environment within the SMR. 

More specifically, as the magnetic field lines thread through the accretion disk, the electric field component $\bm{E}_{\parallel}$ along those lines accelerate charged particles and lift them out of the disk. The charges are subsequently distributed along the magnetic lines in such a way that the lines become equipotential, shutting down further particle extraction and establishing a state of quasi-equilibrium. This enforces $E_i B^i =0$, which is a necessary condition for force-free electrodynamics. The charged particle density is then the minimum value sufficient to short out $\bm{E}_{\parallel}$. 

We now turn to computing the RM-inferred $n_e$ (after excluding the intergalactic and Milky Way contributions), and see if it is indeed consistent with a force-free environment. We start with the following ingredients:
\begin{list}{\labelitemi}{\leftmargin=1em}\renewcommand{\labelenumi}{\textbf{\theenumi}.}
\setlength{\leftmargin}{0pt}

\item 
The magnetic field strength in the SMR is determined by the requirement that it exerts significant influence on the pulsar magnetosphere at its LC, therefore we set its value to $B_{11}$G. 

\item We take the geometric factor from projecting the magnetic field onto our line of sight to be on the order of unity.
We also assume that this line traverses a large portion of the SMR, so that the RM measurement for FRB 110523 is not a chance underestimate of the average FRB value. 

\item For the Milky Way, $\sim 0.01-0.1$pc (scales to other galaxies linearly with SMBH mass) is where the matter is captured by Sgr A* and begins infalling (\cite{2013pss5.book..243M,1994ApJ...426..577M,2002ApJ...575..855Q}), which means that an accreting-material-supported magnetic field extending out to around $L^{\rm SMR} \sim 0.005$pc should be a safe guess (the headroom is larger for heavier SMBHs). 
\end{list}
From these values, the $n_e$ comes out at $0.04/B_{11} {\rm cm}^{-3}$. For a reference scale, we can compute the $n_e$ of the pulsar magnetosphere near its LC (where the magnetic field strength is similar), which is given by 
\cite{Goldreich:1969sb} Eq.~9 (not strictly valid at the LC, but the Lorentz factor drops sharply as we move inward from it, so that this is still an approximate for nearby locations) as
$7\times 10^{-2} B_z/P \sim 0.07 B_{11} \rm{cm}^{-3}$, where $B_z \sim B_{11}$ is the magnetic field strength along the rotation axis in Gauss, and $P\sim 1$s is the pulsar period. 
One major difference exists between the SMR and the pulsar magnetosphere,
however. The positively charged particles in the SMR are protons (with number density $n_p=n_e$) instead of the less massive positrons. Therefore, to ensure self-consistency, we need to check that the energy density of the protons is still subdominant to that of the EM field in the SMR. A simple computation shows that the ratio between them (proton over field) is $\sim 2\times 10^{-3}\gamma B_{11}^{-3}$ , where $\gamma$ is the average Lorentz factor of the protons. Therefore, with a plasma temperature lower than $\sim 10^{15}B^3_{11}$K, our $n_e$ estimate is indeed consistent with the force-free assumption.

\subsection{Dspersion measure and the scattering tail}
\begin{figure}
\begin{overpic}[width=0.99\columnwidth]{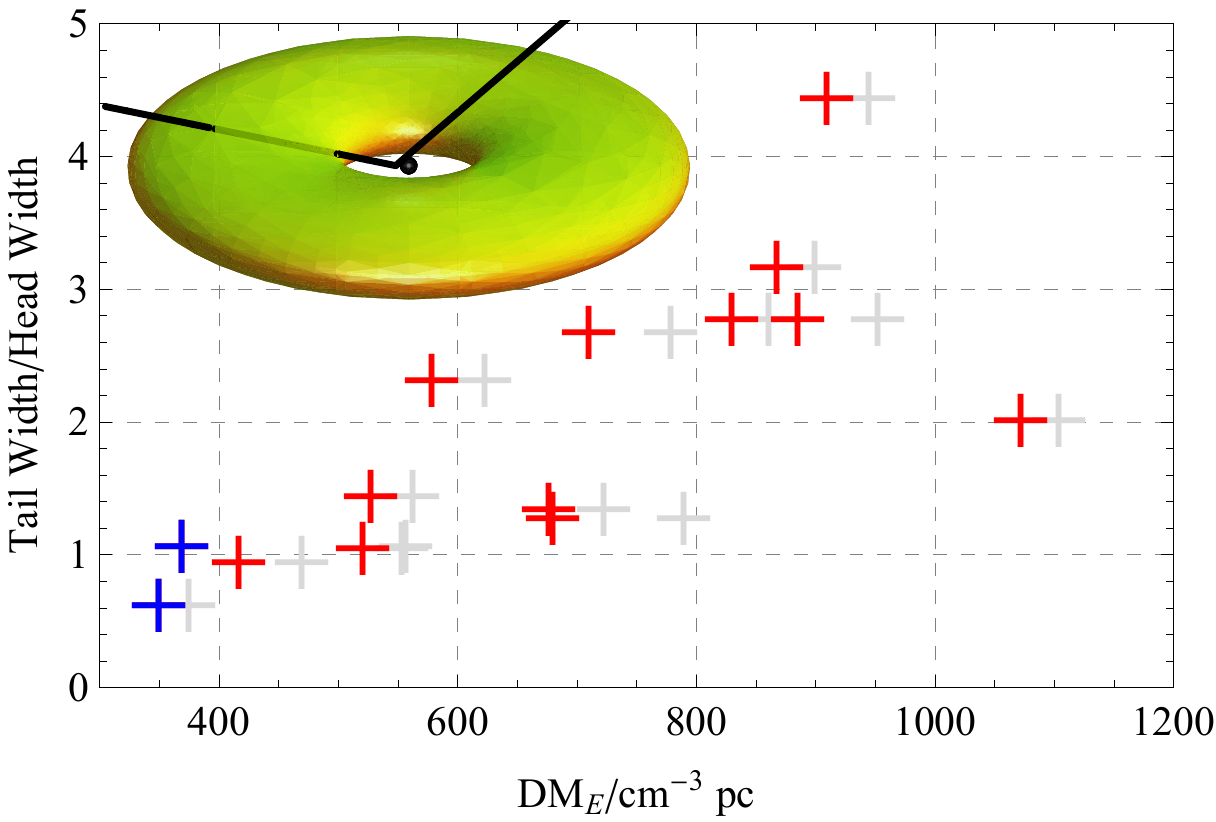}
\put(19,58){I}
\put(43,57){II}
\end{overpic}
\caption{DM against tail-to-head ratio. The tail (head) section width is the distance from the end (beginning) of the pulse to its peak. The red crosses correspond to DM values without the Milky Way contribution (i.e., ${\rm DM}_{E}$), and the gray shadow crosses provide the original overall DM for reference. The two blue crosses correspond to the two pulse profiles plotted in Fig.~\ref{fig:Fitting} and replaces two red crosses. The inset shows the two lines of sight. No points are obscured by the inset. }
\label{fig:DMVsScattering}
\end{figure}

A galactic-center location for FRB sources has been invoked by \cite{2015ApJ...807..179P} to account for the large DMs. The high electron density within the nucleus of the host galaxy is made responsible for the majority of the DM, and the remainder places the sources at extragalactic but non-cosmological distances of hundreds of millions of parsecs. Our scenario is somewhat different, as we place the pulsars close to the SMBH in a relatively clean SMR, in the eye of any scattering screen of a toroidal topology (see the inset of Fig.~\ref{fig:DMVsScattering}). 
Therefore, unless we view the host galaxy approximately edge-on, the radio signal would not have passed through the central core of the screens, and the intergalactic medium would still have contributed a significant fraction to the total DM. (The local SMR contributes a negligible amount to the DM. For example, with FRB 110523, the SMR contribution to DM can be estimated as $\sim {\rm RM}/(0.812 B_{\parallel}) \sim 10^{-4}/B_{11} {\rm cm}^{-3}{\rm pc}$ (note $B_{\parallel}$ is in $\mu$G), out of a total of $\sim 600 {\rm cm}^{-3}{\rm pc}$. The SMR is also not significant in its contribution to the scattering tail, as it is physically too small, thus all locations within are too close to the source).
In fact, the scattering tail may offer a way to distinguish which FRBs are viewed through a screen. 

Some FRBs (e.g.,~FRB 110220) show clear exponential tails typical of multipath broadening, and \cite{2014ApJ...785L..26L} also showed that the intergalactic medium does not appear sufficient to achieve this amount of scattering. Therefore, screens closer to the sources need to be involved (the detection of scintillation in FRB 110523 further supports this scenario).  
On the other hand, some signals, including those plotted in Fig.~\ref{fig:Fitting}, do not display the signature of a scattering tail as clearly. One possibility is that in some cases, our line of sight (line I in the inset of Fig.~\ref{fig:DMVsScattering}) threads through the scattering screens in the galactic center, while in other cases, we peek a more direct view (line II). If this is true, then repeat bursts should retain their classification under the dichotomy, as they originate from the same source, an expectation that appears to have been borne out (\cite{Spitler:2016dmz,Scholz:2016rpt}). 

Moreover, we should expect statistically that signals with larger DMs would exhibit more pronounced scattering signatures, manifesting themselves as a fatter tail when compared to the width of the leading edge (lack of such an asymmetry is quoted in \cite{Spitler:2016dmz} as a sign of the missing scattering tail). In other words, we normalize the tail by the intrinsic width of the signal, instead of assigning all of the extendedness of the trailing half or the whole of the profile as being due to scattering, which may lead to significant overestimates (a simple plot of DM versus pulse width shows no discernible correlation). 
Our investigation is of course further complicated by the fact that we cannot remove the intergalactic component from the overall DM, therefore a correlation will be loose at best, but a scatter plot of DM vs width ratios could still demonstrate a bias provided that the screen contribution to the DM is not completely overwhelmed. We make such a plot in Fig.~\ref{fig:DMVsScattering}\footnote{The data are taken from \cite{2013Sci...341...53T} for FRB 110220, FRB 110627, FRB 110703, FRB 120127; from \cite{2007Sci...318..777L} for FRB 010724; from \cite{2014ApJ...792...19B} for FRB 011025; from \cite{2016MNRAS.tmpL..49C} for FRB 090625, FRB 130626, FRB 130628, FRB 130729; \cite{2015ApJ...799L...5R} for FRB 131104; \cite{Spitler:2016dmz} for FRB 121102; \cite{2015MNRAS.447..246P} for FRB 140514; \cite{2015Natur.528..523M} for FRB 110523. We chose the single-peak profiles, with the highest signal-to-noise ratio when data from multiple bursts were available. In particular, this excludes FRB 121002 from \cite{Thornton13}, which has the largest DM. The temporal sampling rate for this double-peaked signal is relatively low, but the fitted profile does not appear to show a scattering tail.}, which does appear to display a general trend. With better data quality and quantity, and a much more
refined method for extracting the scattering contribution to the trailing edge, 
it may become possible to separate the intergalactic medium contribution to the DM from that of the screen, with the former appearing as a stochastic spread in the horizontal direction around a smooth monotonic curve representing the latter (assuming the existence of a universal temporal scattering versus DM relationship for screens across galaxies). 

\subsection{Event rate} \label{sec:EventRate}
The FRBs have a high implied aggregate event rate of up to $7\times 10^4$ per day at above $1.5$Jy ms fluence across all sky directions (\cite{2015ApJ...807...16L}). There are two ways to achieve a high rate: a large number of independent sources, or large number of repeat events from each individual source. 
The bubble-bursting model falls within the second category. If $500 $ Mpc is to be used as a conservative detectability horizon for the FRBs (the DM is contaminated by the source's local environment, therefore we take predictions from \cite{2015ApJ...807..179P} as a lower bound), 
then there would be millions of galaxies within a volume of that radius, assuming a $\sim 10^{11}$ total galaxy population in the observable universe (\cite{2005ApJ...624..463G}). Furthermore, assuming that there is an SMBH at the center of each, we reach the upper FRB rate limit with an average recurrence rate of once per month 
(lower than the upper limit of $3.2$ per day set by \cite{2015ApJ...807...16L})
if there are $\sim 0.5$ pulsars in the strongly magnetized central region. The number of normal pulsars close to our Sgr A* was estimated by \cite{2004ApJ...615..253P} to be around $100-1000$ within about $0.02$pc from it,
and with a naive volumetric averaging, the number of pulsars expected within a SMR of $L^{\rm SMR}\sim 0.005$pc is $1.6-16$, which covers our required population comfortably. We caution, however, that there appears to be a ``missing-pulsar problem" (\cite{2010ApJ...715..939M,2014ApJ...783L...7D}). 
Within the present model, however, it is relatively easy to accommodate any deficiencies in the galactic center pulsar number by noting that the size of $L^{\rm SMR}$ can be increased further with more massive SMBHs, and that the burst repeat frequency as well as the detectability horizon also have some room to expand into. 

\subsection{Correlated observations}
The observables considered so far have been somewhat circumstantial in terms of placing the sources of the FRBs in galactic centers. In this section, we briefly consider the possibility and difficulties in using correlated observations to carry out more reliable source localizations. 

Although for our proposed FRB mechanism to operate, the pulsars can be on stable orbits around the SMBH and do not need to be falling directly into it, occasions may nevertheless arise when the latter scenario becomes realized, such as when orbital parameters are altered by perturbations by other stars. This may present us with a distinguishing test for the mechanism, namely that the observation of the neutron star's tidal disruption by the SMBH (for example, if the disruption occurs before the NS reaches the innermost stable circular orbit, or ISCO, of the SMBH, the gravitational wave signal detectable by space-based detectors will exhibit a distinct cutoff frequency) should be preceded by a sequence of FRBs from essentially the same location. 

We can quickly assess the feasibility of this particular type of correlated observations. The disruption of the NS by a black hole is highly dependent on the NS equation of state, and even more sensitively on the black hole mass. As we have seen in Sect.~\ref{sec:tidal}, larger black holes tend to lead to less pronounced tidal effects. While SMBHs regularly disrupt normal stars (e.g., Swift J1644+57), they will not be able to disrupt neutron stars outside of the ISCO, regardless of the equation of state (see Table II in \cite{2010PhRvD..81f4026F} for the black hole mass limit beyond which no disruption occurs; they range from several to tens of solar masses for different equations of state, many orders of magnitudes smaller than the masses of SMBHs), and unlikely to do so before the NS enters the event horizon (for extremally spinning black holes in particular, the ISCO is on the horizon, see \cite{ted}). This is unfortunate, but hope remains that binary neutron star or neutron star-stellar mass black hole mergers may occur close to an SMBH, as such regions with dense stellar populations should be hotbeds for dynamical capture, leading to highly eccentric binaries that merge quickly (see \cite{O'Leary:2008xt,Antonini:2012ad}, and also \cite{2009ApJ...700.1933H} for an estimate on binary fraction). When this scenario is realized, a gravitational wave detection would precede and forewarn electromagnetic telescopes, which should subsequently observe, among other things, a short gamma ray burst (SGRB) and an optical kilonova (\cite{metzger:11}) that may allow for identifying the signal as being from a galactic center environment 
(with optimal conditions, even the gravitational wave signal alone may display the effects of a nearly SMBH, see \cite{2012PhRvD..86f4030Z}). Before these observations, however, the NSs should have generated a string of FRB signals that we may try to dig out of archival data. Furthermore, if the merger remnant is another neutron star, then we may have additional FRBs after the merger. By comparing with, for example, SGRB sources outside of galactic centers (should be without correlated FRBs), we may obtain strong indications as to whether FRBs are indeed associated with galactic center locations.

In addition to correlated but separate events, search has also been ongoing for afterglows of the same event that generated FRBs (see, e.g., \cite{Keane:2016yyk}). No confirmed observations have been made (\cite{2016ApJ...821L..22W}), and the present proposal predicts little afterglow. 
To arrive at this prediction, we note that once the closed field lines open up, the stored dipolar energy can stream out along the monopolar field lines as Alfv\'en waves. Analytical solutions for this process exists (see \cite{Brennan:2013jla,2014PhRvD..89j3013B,Zhang:2015aga}), demonstrating that such waves are stable and highly efficient transporters of energy, as they can propagate cleanly without being back-scattered. They would thus evacuate excess energy out of the busted bubble regions quickly, without any residuals hanging back or sloshing around to power long-duration afterglows. 

Nevertheless, temporally coincident observations to FRBs in other electromagnetic frequency bands may be possible. The aforementioned waves can take on the characteristics of a wind (\cite{Brennan:2013jla}). We may therefore reasonably expect that as the pulse of wind slams into the interstellar medium, a ``termination shock" may become visible in a broad frequency range up to $\gamma$-rays, in analogy with pulsar wind nebulae. Coincident detections in multiple bands should at the very least provide additional extinction data, aiding the effort to assess whether the FRBs occur in galactic centers.

\section{Conclusion} \label{sec:Con}
The FRBs are intriguing phenomena, their frequent appearances and consistency (especially on energy scales) indicate a reliable underlying driving mechanism, preferably relying on no incidental environmental factors that can vary significantly from event to event. We proposed a candidate model that attempts to adhere to this observation. We began by noting that the magnetic field strength within galactic nuclei is comparable to the intrinsic pulsar field at its LC, which is a vital but potentially fragile place for the pulsar magnetosphere, harboring such extreme conditions as particles traveling at the speed of light.
We then showed that the galactic magnetic field exerts a destabilizing influence there by imposing discontinuous Lorentz forces on sheets of high surface current density that enclose dipole-like closed magnetic field lines. The result is that the CSs collapse and the closed lines open up into a split-monopole configuration, with reduced field strengths so that they can retain the same asymptotic boundary conditions farther away from the star. This implies a reduction of energy density in the magnetosphere, with the difference becoming available to fuel FRBs. The FRB energy budget thus computed agrees with observed values, and several other aspects of the process also appear to be compatible with observations. 

We note, however, that although we have sketched the aforementioned synopsis, a detailed blueprint of the dynamical energy-release process requires sophisticated numerical simulations, where a realistic treatment of the resistivity arising from turbulences excited during the bubble-bursting process would be essential. In short, the most important magnetospheric dynamics involved are those for which the force-free assumption does not apply. Unfortunately, these are also the processes that we understand the least. With this paper, we wish to highlight this issue and evoke further discussions, especially on the dynamical evolution of the CSs.

\begin{acknowledgements}
We thank an anonymous referee for insightful comments and suggestions that led to the introduction of Sects. 5,  6, and 7.5.
This work is supported by the NSFC Grant 11503003, the Fundamental Research Funds for the Central Universities Grant 2015KJJCB06, and a Returned Overseas Chinese Scholars Foundation grant. 
\end{acknowledgements}

\bibliographystyle{aa} 
\bibliography{paper.bbl} 

\end{document}